\documentclass[prb,showpacs,floatfix,superscriptaddress,amsmath,amssymb,twocolumn]{revtex4}
\usepackage{graphicx}
\usepackage{amssymb}
\usepackage{amsmath}
\usepackage{color}
\usepackage[colorlinks,bookmarks=false,citecolor=blue,linkcolor=red,urlcolor=blue]{hyperref}

\newcommand\be{\begin{equation}}
\newcommand\ee{\end{equation}}

\begin{document}

\title{Parametric instability in periodically driven Luttinger Liquids}

\author{M. Bukov}
\affiliation{Department of Physics, Arnold Sommerfeld Center for Theoretical Physics,
and Center for NanoScience, Ludwig-Maximilians-Universit\"at M\"unchen,
Theresienstr. 37, 80333 Munich, Germany}
\author{M. Heyl}
\affiliation{Department of Physics, Arnold Sommerfeld Center for Theoretical Physics,
and Center for NanoScience, Ludwig-Maximilians-Universit\"at M\"unchen,
Theresienstr. 37, 80333 Munich, Germany}
\affiliation{Institut f\"ur Theoretische Physik, Technische Universit\"at Dresden, 01062 Dresden, Germany}

\begin{abstract}
	We analyze the properties of a Luttinger liquid under the influence of a periodic driving of the interaction strength. Irrespective of the details the driven system develops an instability due to a parametric resonance. For slow and fast driving, however, we identify intermediate long-lived meta-stable states at constant time-averaged internal energies. Due to the instability perturbations in the fermionic density are amplified exponentially leading to the buildup of a superlattice. The momentum distribution develops a terrace structure due to scattering processes that can be associated with the absorption of quanta of the driving frequency.
\end{abstract}


\maketitle

\section{Introduction}

Due to remarkable progress in experiments it is within the scope of present technology to implement and simulate the dynamics of quantum many-body systems with a high degree of controllability on the system parameters even under nonequilibrium conditions. While Quantum Dots provide the framework for the study of quantum impurity systems~\cite{Exp_Kondo}, ultra-cold atoms in optical lattices constitute the basic structure for the experimental realization of interacting quantum many-body systems such as the Bose- or Fermi-Hubbard model~\cite{Optical_lattices}.

In principle, one can imagine a variety of different nonequilibrium driving protocols. By now the nonequilibrium dynamics following interaction quenches or ramps is studied in great detail~\cite{Coll_quench}. The influence of a periodic driving, however, on strongly correlated many-particle systems poses new challenges especially in terms of methodology and has only been studied for a restricted class of systems. There has been considerable work on periodically driven strongly correlated impurity systems such as Anderson impurity~\cite{AIM_per} and Kondo models~\cite{Heyl_Kondo,Kondo_per}. Besides impurity systems, Falicov-Kimball~\cite{Fal_Kim,Fal_Kim_diss} and Hubbard models~\cite{Hubbard_DMFT, Hubbard_diss} have been analyzed on the basis of a nonequilibrium extension of dynamical mean-field theory. Periodically driven systems of interacting fermions in one dimensions have been studied for small system sizes~\cite{Hubbard_1D} and in the Luttinger liquid limit~\cite{Kagan,Graf,Pielawa}. Periodically driven one-dimensional Bose-Hubbard models have been investigated by exact diagonalization for small systems~\cite{Holthaus} and for large systems based on the time-dependent density matrix renormalization group~\cite{Poletti}.

In this work we investigate the nonequilibrium dynamics of interacting fermions in one dimension within a Luttinger liquid description induced by a periodic time-dependence of the interaction strength. The impact of a periodic modulation of the Fermi velocity onto the fermionic momentum distribution has been investigated recently~\cite{Graf}. Due to the above mentioned complexity of periodically driven systems it is instructive to explore those particular cases where exact and nonperturbative solutions are accessible such as in the case studied in this work.

The periodically driven Luttinger liquid shows an instability in the long-time limit due to a parametric resonance~\cite{Kagan,Graf,Pielawa}. However, on intermediate time scales meta-stable steady states of constant time-averaged energy densities can form with long lifetimes in case of slow and fast driving. In the adiabatic limit the instability occurs after a fixed number of periods independent of the driving frequency. Thus it is impossible to adiabatically follow the ground state of the system for a large number of periods irrespective of how slowly the system is driven. The periodic driving promotes fermionic scattering processes under the absorption of quanta of the driving frequency yielding a terrace structure in the fermionic momentum distribution. In the slow driving limit the system becomes unstable against perturbations in the fermionic particle density resulting in an exponential amplification of the perturbation in consequence of the parametric resonance.

This paper is organized as follows. In Sec.~\ref{sec:ll} we introduce the model of a periodically driven Luttinger liquid whose dynamics including the parametric instability we analyze in Sec.~\ref{sec:neq_evol}. In Sec.~\ref{sec:energy} we identify a meta-stable state for fast and slow driving based on the study of the internal energy density. The impact of the instability on the fermionic density and the momentum distribution is analyzed in Sec.~\ref{sec:density} and Sec.~\ref{sec:momentum}, respectively.

\section{Periodically driven Luttinger liquid}
\label{sec:ll}

Consider a system of 1D fermions of length $L$ whose interaction strength is varied periodically with a frequency $\Omega$ termed the driving frequency in the following. Introducing left- as well as right-moving fermions indicated by a label $L/R$ and linearizing the dispersion relation around the respective Fermi points one arrives at the following Hamiltonian:
\begin{eqnarray}
	H & = & H_0 + H_{int} \label{eq:LL_Hamiltonian} , \\ 
		H_0 &=& v_F \int \frac{dx}{2\pi} \colon \left[ \psi_L^\dag(x) i \partial_x \psi_L(x) - \psi_R^\dag(x) i \partial_x \psi_R(x)  \right] \colon, \nonumber \\
		H_{int}& = & \sum_{\eta,\eta'}\int \frac{dx}{2\pi} \frac{dx'}{2\pi}  \colon \rho_\eta(x) \colon \frac{1}{2} U(x-x';t) \colon \rho_{\eta'}(x') \colon. \nonumber
\end{eqnarray}
This Hamiltonian differs from the equilibrium case only through the periodic time dependence of the interaction $U(x-x';t)$ with an associated Fourier transform $ U_q(t) = \int dx \, e^{-iqx} \, U(x;t)$. We assume a repulsive interaction potential $U_q(t)>0$ of finite range such that it is cut off beyond some momentum scale $q_c$. The colons $\colon \dots \colon$ denote normal ordering relative to the Fermi sea and $v_F$ is the Fermi velocity. The fermionic density $\rho_\eta(x)$ with $\eta=L/R$ is determined by the fermionic fields $\psi_\eta(x)=\sqrt{2\pi/ L} \sum_k e^{-ikx} c_{k\eta}$ via $\rho_\eta(x) =\colon \psi_\eta^\dag(x) \psi_\eta(x) \colon $. The operator $c_{k\eta}^\dag$ creates a fermion of the species $\eta=L/R$ with wave vector $k$ .

Note that we restrict to the case of spinless fermions. This reduced model system already incorporates most of the characteristic features of interacting fermions in 1D. For a system of fermions with spin in equilibrium, for example, the dynamics separates into two independent sectors of spin and charge, a phenomenon called spin-charge separation, each of which can be modeled by a Hamiltonian of the form in Eq.~(\ref{eq:LL_Hamiltonian}). Note that the influence of a periodic modulation of the Fermi velocity onto the momentum distribution in case of fermions with spin has been investigated recently~\cite{Graf}.

In the nonequilibrium scenario under investigation the system is initially prepared in the ground state $|\psi_0\rangle$ at some fixed interaction strength such that $U_q(t<0)=V_q(1+\nu)$ is chosen to be time-independent for times $t<0$. At time $t=0$ the periodic driving is started with the following parametrization of the time-dependence of the interaction potential
\be
	U_q(t) = V_q \left( 1 + \nu \cos(\Omega t) \right).
\label{eq:potential}
\ee
The dimensionless coupling $\nu$ of the periodic driving is chosen $\nu<1$ such that the interaction remains repulsive for all times. The interaction strength is characterized by the dimensionless number
\be
	\alpha = \frac{V_0}{2\pi v_F}
\ee
which within the validity of the Luttinger model is always chosen $\alpha<1$. We assume that the interaction potential $V_q$ is cut off beyond the momentum scale $q_c$. In our numerical simulations we choose a Gaussian for simplicity, i.e., $V_q/(2\pi v_F) = \alpha \exp[-(q/q_c)^2]$.

Although the Hamiltonian in Eq.~(\ref{eq:LL_Hamiltonian}) is quartic in fermionic operators it can be mapped onto a quadratic and exactly solvable problem using the bosonization technique~\cite{Schoeller_Delft}. Introducing bosonic operators~\cite{Schoeller_Delft}
\begin{align}
	& b_{q\eta} = -i \sqrt{\frac{2\pi}{Lq}} \sum_k c_{k-q \eta}^\dag c_{k\eta}, \:\: q>0, \nonumber\\
	& b_{q\eta}^\dag = i \sqrt{\frac{2\pi}{Lq}} \sum_k c_{k+q \eta}^\dag c_{k\eta}, \:\: q>0,
\end{align}
for each right- and left-moving branch $\eta=L/R$, the Hamiltonian in Eq.~(\ref{eq:LL_Hamiltonian}) can be mapped onto a quadratic but time-dependent bosonic problem
\begin{align}
	H & = \sum_{q>0,\eta=L/R} \omega_q(t) b_{q\eta}^\dag b_{q\eta} - \\ & - \sum_{q>0} q \frac{U_q(t)}{2\pi} \left[ b_{qL}^\dag b_{qR}^\dag + b_{qR}b_{qL} \right] + \Delta(t).
\label{eq:LL_bosons}
\end{align}
The dispersion of the diagonal part of the above Hamiltonian is given by
\be
	\omega_q(t) = q v_F \left( 1 + \frac{U_q(t)}{2\pi v_F} \right).
\label{eq:omegaq}
\ee
The overall constant $\Delta(t)= (2\pi)^{-1}\sum_{q>0} q U_q(t)$ has no effect on the time evolution of observables except the internal energy itself as discussed in Sec.~\ref{sec:energy}. In principle, it is possible to diagonalize this Hamiltonian using a time-dependent unitary transformation. However, it turns out to be suitable to determine the dynamics in the untransformed basis, see Sec.~\ref{sec:neq_evol} below.

\section{Nonequilibrium time evolution}
\label{sec:neq_evol}

For the dynamics of all quantities considered such as energy density, fermionic density, and the momentum distribution it is sufficient to solve the Heisenberg equations of motion for the bosonic operators
\be
	\frac{d}{dt} b_{q\eta}(t) = -i\omega_q(t) b_{q\eta}(t) + iq\frac{U_q(t)}{2\pi}b_{q \overline{\eta}}^\dag 
\label{eq:eom}
\ee
with $\overline{\eta}$ the conjugate species of $\eta$, i.e., $\overline{L}=R$ and vice versa. These differential equations for operators can be transformed into differential equations for complex functions $\chi_{q\eta}(t)$ and $\lambda_{q\eta}(t)$ defined by
\be
	b_{q\eta}(t) = \chi_{q\eta}(t) b_{q\eta} + \lambda_{q\eta}(t) b_{q\overline{\eta}}^\dag.
\ee
when inserted into Eq.~(\ref{eq:eom}). The resulting system of coupled differential equations can be cast into a more familiar form by regarding appropriate superpositions
\be
	\alpha_{q\eta} = \chi_{q\eta} - \lambda_{q\eta}^\ast,\:\: \beta_{q\eta} = \chi_{q\eta} + \lambda_{q\eta}^\ast.
\ee
The function $\alpha_{q\eta}$ is the solution of a parametrically driven harmonic oscillator and obeys a Mathieu equation in properly scaled parameters
\be
	\frac{d^2\alpha_{q\eta}(\tau)}{d\tau^2} +  \epsilon_q^2\left[1 + 2\gamma_q \cos(2\tau) \right] \alpha_{q\eta}(\tau) = 0
\label{eq:Mathieu}
\ee
with the dimensionless time $\tau = \Omega t/2$, the natural frequency $\epsilon_q=2 v_F q \Omega^{-1}\sqrt{1+V_q/(\pi v_F)}$ of the harmonic oscillator, and $\gamma_q = \nu V_q/(2\pi v_F+2V_q)$ the coupling strength of the periodic perturbation. The initial conditions for the solution of the Mathieu equation are $\alpha_{q\eta}(t=0) = 1$ and $\alpha_{q\eta}'(t=0)=-iv_Fq$. The remaining function $\beta_{q\eta}=i(qv_F)^{-1} d\alpha_{q\eta}(t)/dt$ is proportional to the time derivative of $\alpha_{q\eta}$.

Concluding, the time evolution in the periodically driven Luttinger liquid is equivalent to a set of parametrically driven harmonic oscillators. The Mathieu equation in Eq.~(\ref{eq:Mathieu}) in general exhibits no analytic solution in terms of elementary functions. For special cases, however, such as parametric resonance, approximate analytical solutions are available, see below. For the general case we solve the differential equations numerically using a standard 4-th order Runge-Kutta algorithm.

The driven harmonic oscillator in Eq.~(\ref{eq:Mathieu}) shows an instability with exponentially growing amplitudes in the case of parametric resonance which occurs for that particular plasmonic mode $q^\ast$ for which $\epsilon_{q^\ast}=1$ or equivalently $\Omega/2 = v_F q^\ast \sqrt{1+V_{q^\ast}/(\pi v_F)}$~\cite{Landau}. 

In the following it is important to distinguish two different cases of fast and slow driving. The energy scale $\Omega^\ast$ associated with the crossover between the two limits is set by
\be
	\Omega^\ast = v_F q_c.
\ee
For slow driving $\Omega \ll \Omega^\ast$ the resonant bosonic mode $q^\ast$ for which parametric resonance occurs is determined by
\be
	\frac{q^\ast}{q_c} \stackrel{\Omega \ll \Omega^\ast}{\longrightarrow}\frac{1}{\sqrt{1+2\alpha}}\frac{\Omega}{2\Omega^\ast}
\label{eq:qast_slow}
\ee
to leading order in $\Omega/\Omega^\ast$. Note, however, that not only $q^\ast$ but also momenta $q$ within a finite interval of nonzero length contribute to the resonance~\cite{Landau}. The rate $\Gamma$ of the associated exponential growth in time can be determined using standard methods~\cite{Landau}
\be
	\Gamma \stackrel{\Omega\ll\Omega^\ast}{=} \frac{1}{4} \frac{\alpha \nu}{1+2\alpha}\Omega.
\label{eq:Gamma_slow}
\ee
The time scale $t^*$ for the onset of the instability is then determined by the rate $\Gamma$ via
\be
	t^*=\Gamma^{-1}.
\label{eq:tstar}
\ee
In the opposite case $\Omega \gg \Omega^\ast$ of fast driving the resonant mode
\be
	\frac{q^\ast}{q_c} \stackrel{\Omega \gg \Omega^\ast}{\longrightarrow} \frac{\Omega}{2\Omega^\ast}
\label{eq:qast_fast}
\ee
is independent of the interaction potential up to corrections suppressed by the cutoff $q_c$. The associated rate of the exponential growth is then given by
\be
	\Gamma \stackrel{\Omega\gg\Omega^\ast}{=} \frac{1}{4} \Omega \nu V_{q^\ast}.
\label{eq:Gamma_fast}
\ee
Its precise behavior for $\Omega/\Omega^\ast \gg 1$ or equivalently $q^\ast/q_c \gg1$ depends on the details of the large momentum behavior of the interaction potential. If $V_q \sim \exp[- C q]$ for some constant $C>0$ the rate $\Gamma \sim \exp[-D\Omega/\Omega^\ast]$ with $D=C q_c/2$ is suppressed exponentially. Analogously, algebraically decaying potentials $V_q \sim (q/q_c)^{-\mu}$ yield a power law dependence $\Gamma \sim (\Omega/\Omega^\ast)^{1-\mu}$ for $\mu>1$.

As $\Gamma$ is linear in $\Omega$ for slow driving, see Eq.~(\ref{eq:Gamma_slow}), the adiabatic approximation breaks down after a fixed number of periods $N_{\mathrm{per}}\sim \Omega/\Gamma =4 (1+2\alpha)/(\alpha \nu)$ set by the interaction strength $\alpha$ and the coupling to the periodic perturbation $\nu$ irrespective of $\Omega$. By reducing the driving frequency one cannot increase the number of periods for the validity of the adiabatic approximation.

\section{Internal energy density}
\label{sec:energy}

Typically, the periodically driven quantum many-body systems considered so far in the literature ignore the possible influence of dissipation mechanisms onto the dynamics, with the exceptions of Refs.~\cite{Fal_Kim_diss, Hubbard_diss}. This is an important issue because the energy in the system will in general increase during the considered nonequilibrium protocol. For bounded Hamiltonians, fermionic or spin systems on a lattice, for example, the internal energy will necessarily saturate, for unbounded Hamiltonians this need not be the case. Even though the internal energy for bounded Hamiltonians will always stay finite, the question whether the unavoidable presence of dissipation mechanisms may at some point in time have a considerable influence onto the dynamics is still largely unanswered. Due to the parametric instability the internal energy of the Luttinger liquid  diverges as we will show below. The existence of the instability naturally sets the time scale $t^\ast$, see Eq.~(\ref{eq:tstar}), beyond which additional internal properties such as the curvature of the fermionic dispersion relation or external dissipation mechanisms have to be included for a realistic description. 

The internal energy density $\mathcal{E}(t)=L^{-1} \langle \psi_0(t) | H(t) | \psi_0(t)\rangle$ of the periodically driven Luttinger liquid system at time $t$ is given by
\be
	\mathcal{E}(t) =  \int_0^\infty \frac{dq}{2\pi} \left[2 \omega_q(t) K_q^{(1)}(t) - q \frac{U_q(t)}{2\pi} K_q^{(2)}(t) \right] + \frac{\Delta(t)}{L}
\label{eq:internal_energy}
\ee
where
\begin{eqnarray}
	K_q^{(1)}(t) & = & \sinh^2(\theta_q) + |\lambda_{qL}(t)|^2 \cosh(2\theta_q) + \nonumber \\ & & + \, \mathrm{Re} [\chi_{qL}(t) \lambda_{qL}^*(t)]\sinh(2\theta_q), \nonumber \\
	K_q^{(2)}(t) & = & \sinh(2\theta_q)\, \mathrm{Re}[\lambda_{qL}^2 + \chi_{qL}^2] + \nonumber \\
	& & + 2 \cosh(2\theta_q) \, \mathrm{Re}[\lambda_{qL}\chi_{qL}],
\end{eqnarray}
and $\omega_q(t)$ given by Eq.~(\ref{eq:omegaq}). The Bogoliubov angles $\theta_q$ for the initial state are determined by the formula $\tanh(2\theta_q) = V_q (1+\nu)/(2\pi v_F + V_q(1+\nu))$. Note that the Luttinger liquid Hamiltonian in Eq.~(\ref{eq:LL_Hamiltonian}) appears typically as the low-energy theory derived from more complicated many-body systems. Within such a mapping additional time-dependent contributions to $\Delta(t)$ in Eq.~(\ref{eq:LL_bosons}) can be generated in the periodically driven case. Those contributions depend on the details of the model and have to be worked out for each particular case. In the present work we are interested in the generic low-energy properties that are all contained in the Luttinger liquid description of the model Hamiltonian in Eq.~(\ref{eq:LL_Hamiltonian}). Thereby we ignore additional contributions to the energy density that are generated by the mapping onto this low-energy theory.

\begin{figure}
	\includegraphics[width=\linewidth]{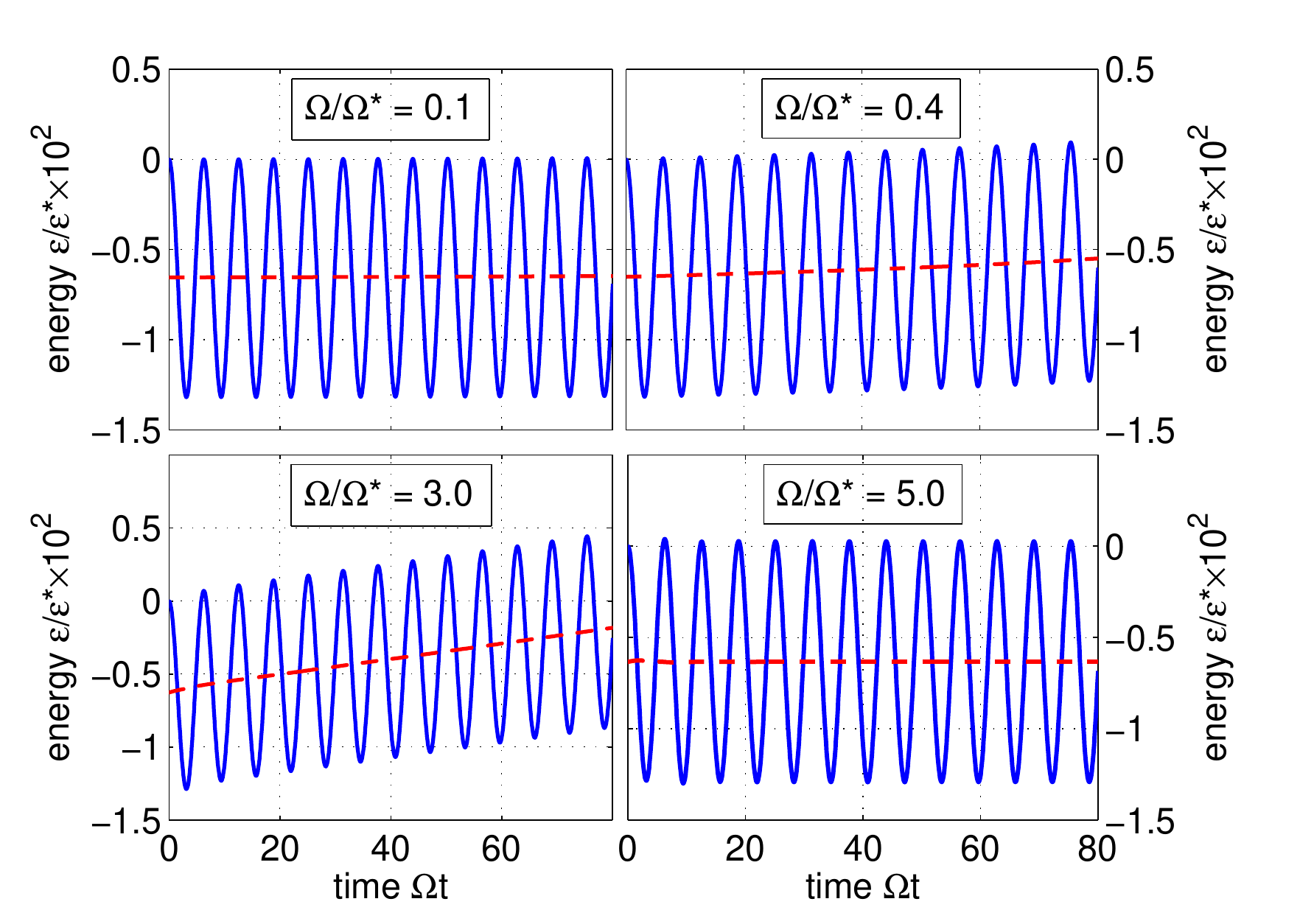}
\caption{(color online) Time evolution of the internal energy density $\mathcal{E}$(t) (blue lines) and its time average $\overline{\mathcal{E}(t)}$ (red dashed lines) for different driving frequencies $\Omega$ in units $\mathcal{E}^\ast = q_c \Omega^\ast$. For these numerical simulations we have chosen an analytic interaction potential $V_q/(2\pi v_F)= \alpha \exp(-(q/q_c)^2)$ with $\alpha = 1/2$ and $\nu=1/5$. The zero of energy has been chosen such that $\mathcal{E}(t=0)=0$.}
\label{Fig1}
\end{figure}

In Fig.~\ref{Fig1} plots for $\mathcal{E}(t)$ are shown for different driving frequencies. Periodically driven systems without instabilities develop stationary states at long times where expectation values time-averaged over one period become time-independent. Besides the internal energy density we have included its time average 
\be
	\overline{\mathcal{E}}(t) = \frac{\Omega}{2 \pi} \int_{t}^{t+2\pi/\Omega} dt' \: \mathcal{E}(t')
\ee
in Fig.~\ref{Fig1} for the identification of such stationary states.

In case of fast driving $\Omega \gg \Omega_\ast$ the rate $\Gamma$ of the parametric resonance is strongly suppressed due to the finite cutoff scale $q_c$ of the interaction potential, see Eq.~(\ref{eq:Gamma_fast}). After the initial transient dynamics following the start of the periodic driving the system settles to a stationary state with a constant time-averaged energy density. As $\Gamma$ is small but still finite the system will nevertheless develop the instability for times $t>t^\ast$. Consequently this intermediate state is only meta-stable.

In the limit $\Omega \to \infty$ the dynamics becomes effectively equivalent to one of a time-averaged Hamiltonian if there exists a mechanism that prevents the absorption of high-energy photons~\cite{Heyl_Kondo}. In the present model system this mechanism is provided due to the finite range $q_c$ of the interaction. Physically speaking, the system is not able to follow the fast external perturbation and only perceives its average contribution. As the initial state is not an eigenstate of the time-averaged Hamiltonian for times $t>0$ the dynamics becomes equivalent to that of an interaction quench. Indeed, we find that the time-averaged energy density $\overline{\mathcal{E}}(t)$ follows precisely the behavior of the interaction quench scenario.

When the driving frequency is lowered the rate of the instability $\Gamma$ grows and thus the onset of the instability moves to smaller times $t^\ast$. For $\Omega \approx \Omega^\ast$ the transient dynamics is directly followed by an exponentially increasing contribution due to the influence of the instability. In this way the instability is so strong that it hinders the buildup of a meta-stable state completely. For even lower frequencies, however, the rate $\Gamma \sim \Omega$ decreases again opening up the window for the meta-stable state as can be clearly observed in Fig.~\ref{Fig1}.

\section{Fermionic density}
\label{sec:density}

The internal energy density of the system mirrors the instability by showing a divergence at times $t>t^\ast$. The question, however, which internal perturbations can prevent the buildup of the instability we have not touched up to now. Regarding the time evolution of the fermionic density we argue that it may be sufficient to include the curvature of the fermionic dispersion relation.
\begin{figure}
	\includegraphics[width=0.9\linewidth]{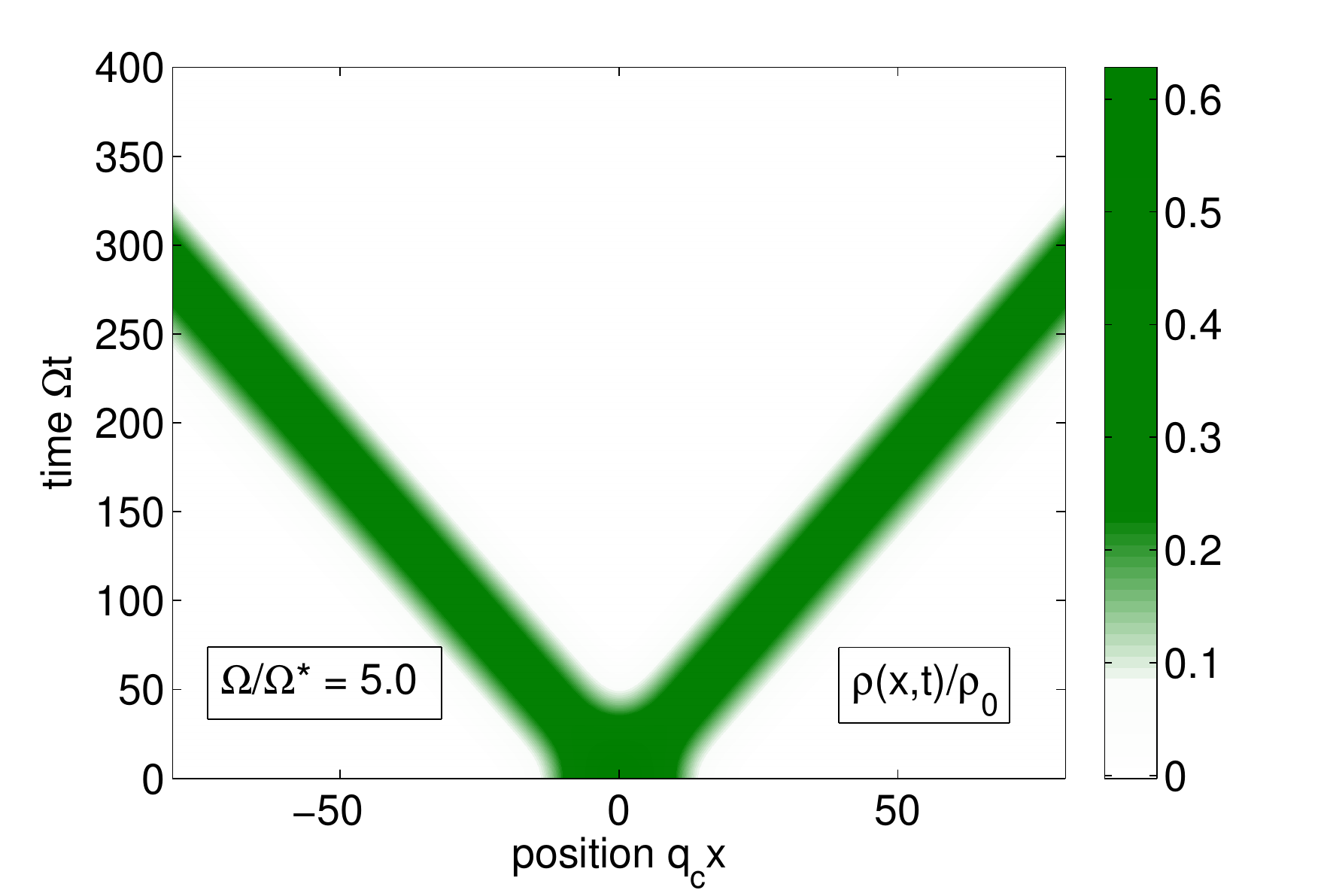} \\ ~\\ ~\\
	\includegraphics[width=0.9\linewidth]{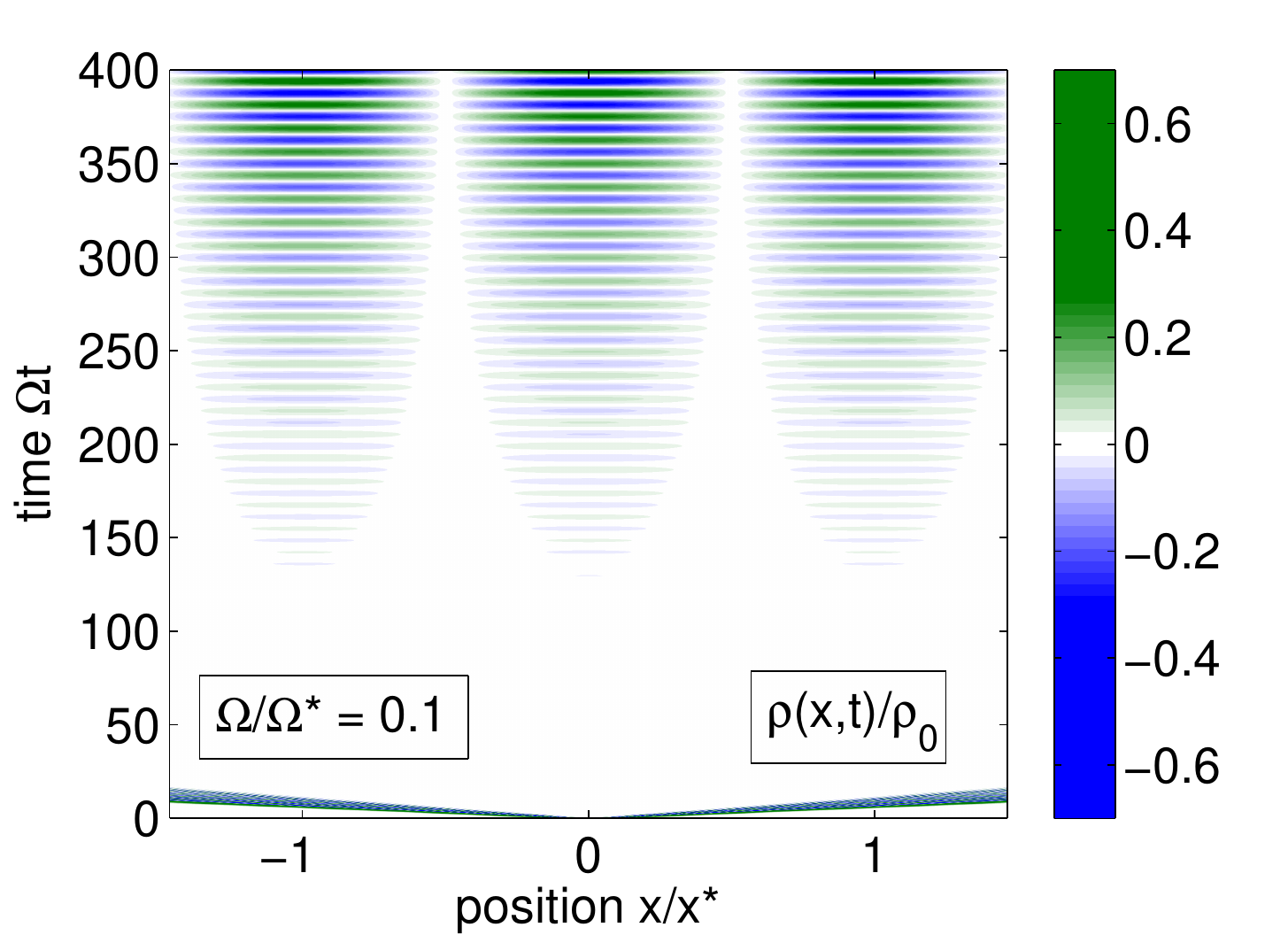}
\caption{(Color online) Time evolution a localized perturbation in the fermionic density $\rho(x,t)$ for the cases of fast (upper plot) and slow (lower plot) driving. Again, we have chosen an analytic interaction potential $V_q/(2\pi v_F)= \alpha \exp(-(q/q_c)^2)$ with $\alpha = 1/2$ and $\nu=1/5$ for these numerical simulations. The initial density profile is a Gaussian $\rho(x,t=0)=\rho_0 \exp(-(x/x_c)^2)/2$ with $x_c q_c=2$ in the slowly and $x_c q_c = 10$ in the fast driven case. The buildup of the superlattice in the slowly driven limit is independent of the precise choice of the width $x_c$ of the initial density wave packet. In the fast driving case, that is equivalent to an interaction quench, see main text, the universal low-energy limit corresponds to $x_c q_c \gg 1$.}
\label{Fig2}
\end{figure}

In the following we analyze the dynamics of an initially localized fermionic density wave packet in presence of the periodic driving. The time evolution of a local perturbation in the fermionic density is solely determined by the solution $\alpha_{q\eta}(t)$ of the Mathieu equation, see Eq.~(\ref{eq:Mathieu}),
\be
	\rho(x,t) = 2 \int_0^\infty \frac{dq}{2\pi} \, \cos(qx) \, \rho_q^0 \, \mathrm{Re} [\alpha_{qL}(t)] 
\label{eq:fermionic_density}
\ee
with $\rho_q^0=\int dq e^{-iqx} \rho_0(x)$ characterising the initial density profile $\rho_0(x)=\rho(x,t=0)$. In Fig.~\ref{Fig2} we show for the fast and slow driving regimes the dynamics of the fermionic density where we have chosen a Gaussian wave packet as initial condition for illustration.

For fast driving $\Omega \gg \Omega_\ast$ the initial local perturbation splits into right- and left-moving contributions. As in case of the internal energy density in Sec.~\ref{sec:density} the dynamics becomes equivalent to the interaction quench scenario. This picture is suitable for times $t<t^\ast$ before the onset of the instability. In this regime the time scale $t^\ast \propto \Omega^{-1}$, see Eq.~(\ref{eq:Gamma_slow}), is large and grows in a power-law fashion for small values of the driving frequency.

In the opposite case $\Omega \ll \Omega_\ast$ a completely different picture emerges. After a fixed number of periods $N_\textrm{per}\sim \Omega/\Gamma = 4(1+2\alpha)/(\alpha \nu) $ to leading order in $\Omega/\Omega^\ast$ the dynamics is dominated by the parametric resonance leading to an exponential growth of the initial perturbation for $x\ll v_{max}t$
\be
	\rho(x,t) \stackrel{t \gg t^\ast}{\longrightarrow} q_c A \cos(q^\ast x)\cos\left( \frac{\Omega t}{2} +\pi/4 \right) e^{\Gamma t}
\ee
with $A$ a constant nonuniversal prefactor. A superlattice forms with a period
\be
	x^\ast=2\pi/q^\ast
\label{eq:xstar}
\ee
set by the resonant mode, see Eqs.~(\ref{eq:qast_slow},\ref{eq:qast_fast}), whose amplitude is growing exponentially at a rate $\Gamma$. Note that this does not lead to a violation of particle number conservation as it might seem from the plot in Fig.~\ref{Fig3}. For each density hump there also exists a valley of depletion of fermionic charge carriers. Moreover, the superlattice extends only over a distance $d\sim v_{max}t$ within the light cone set by the maximal sound velocity. Integrating over the whole  real-space shows that the particle number is still conserved as one can directly check via Eq.~(\ref{eq:fermionic_density}).

Now we want to argue that including the curvature of the fermionic dispersion relation will cut off the exponential growth of the superlattice for sufficiently large densities. In the limit of slow driving we can approximately neglect the influence of the finite range of the interaction and set $V_q \approx V_0$. For a $q$-independent interaction the influence of a nonlinear fermionic dispersion relation can be accounted for approximately~\cite{Rozhkov_Imambekov}. The time-independent version of the Hamiltonian in Eq.~(\ref{eq:LL_Hamiltonian}) including the quadratic curvature contribution can be mapped to a free Fermi gas~\cite{Rozhkov_Imambekov}. For a free Fermi gas the buildup of large densities is prevented due to the nonlinear dispersion eventually leading to the production of shock waves~\cite{Bettelheim}.

\section{Momentum distribution}
\label{sec:momentum}

In equilibrium the momentum distribution for the fermionic particles exemplifies the different influence of repulsive interactions in one dimension compared to higher dimensions where Fermi liquid theory holds. In Luttinger liquids the momentum distribution shows no jump at the Fermi energy even at zero temperature reminiscent of the absence of a finite quasiparticle weight.
\begin{figure}
	\includegraphics[width=1\linewidth]{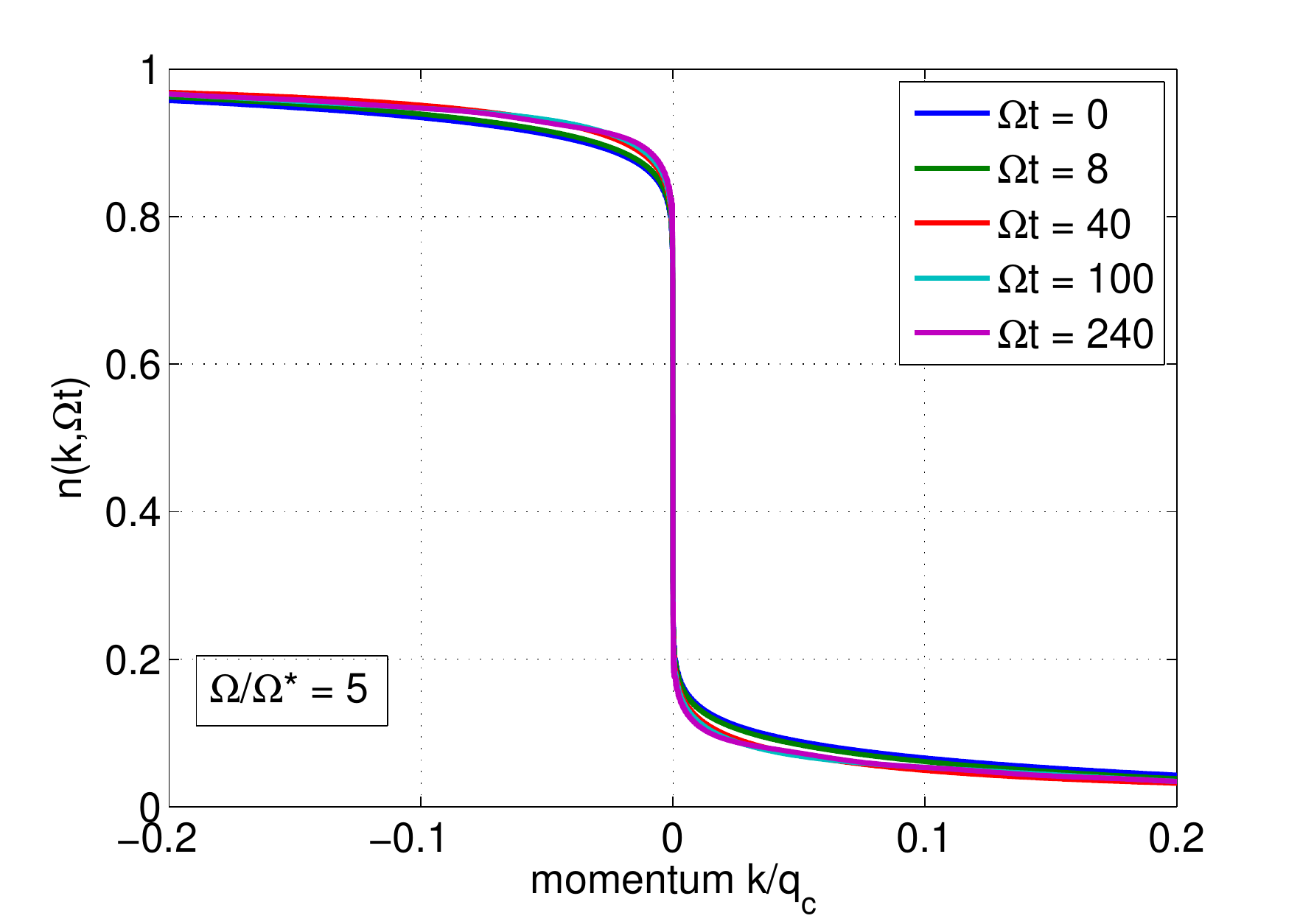} \\~\\~\\
	\includegraphics[width=1\linewidth]{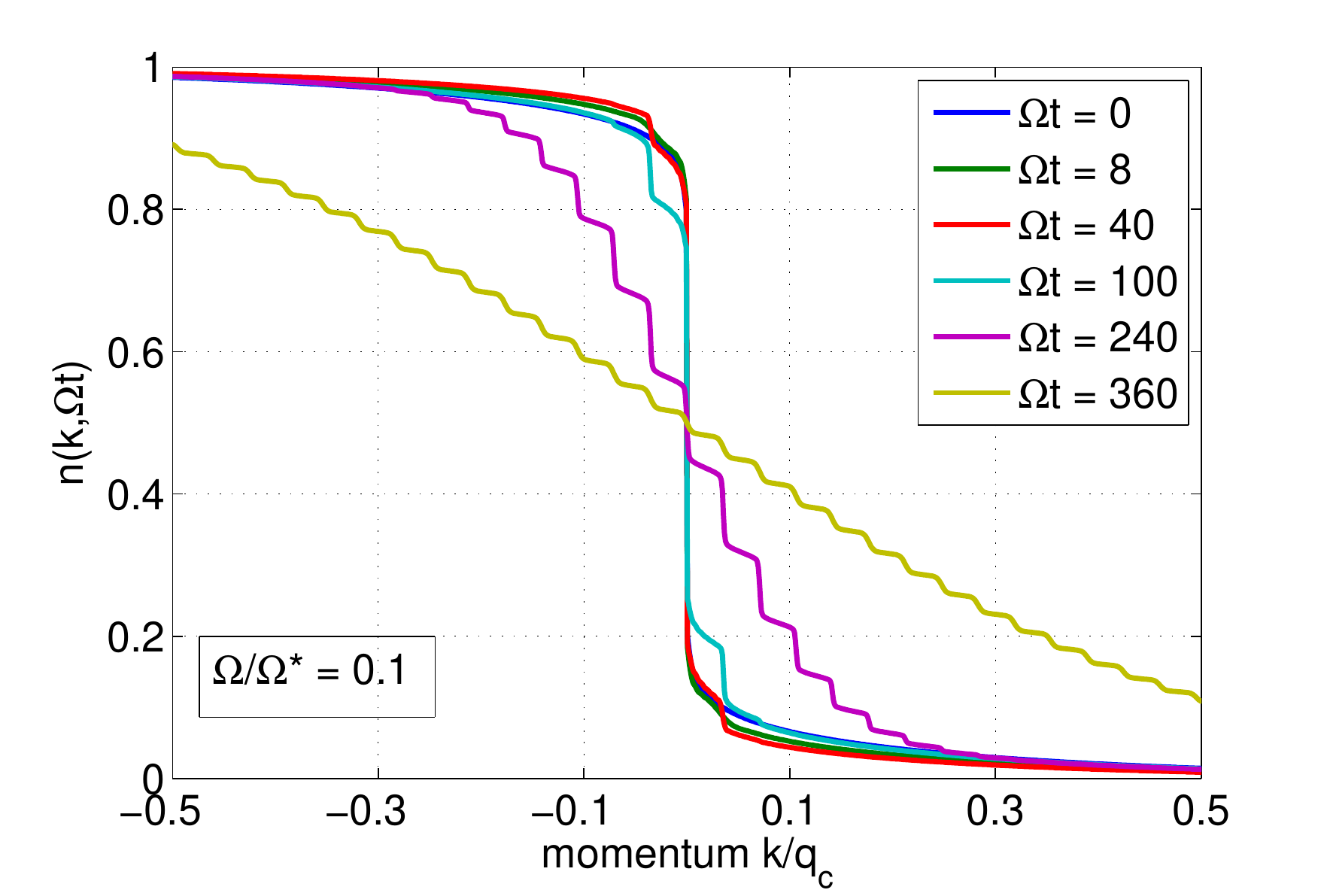}
\caption{(Color online) Time evolution of the momentum distribution $f_k(t)$ for fast (upper plot) and slow (lower plot) driving. The parameters of the numerical simulation are chosen as for the previous plots.}
\label{Fig3}
\end{figure}

The momentum distribution function $f_k(t)$ for the left-moving fermions
\be
	f_k(t) = \langle \psi_0 | c_{kL}^\dag(t) c_{kL}(t) | \psi_0 \rangle, 
\ee
with $|\psi_0\rangle$ the initial state, is connected to an equal-time correlation function $f(x,t)$ in real-space via Fourier transformation
\begin{eqnarray}
	&& f_k(t) =  \int \frac{dx}{2\pi} e^{-ikx} f(x,t), \\
	&& f(x,t) = \langle \psi_0 | \psi_L^\dag(x,t) \psi_L(0,t) | \psi_0\rangle
\end{eqnarray}
that can be calculated analytically using the bosonization technique~\cite{Schoeller_Delft}
\be
	f(x,t) = \frac{1}{a+ix} e^{-F(x,t)}.
\ee
with $a^{-1}$ an ultraviolett cut-off. The rate function $F(x,t)=F_{\textrm{eq}}(x) + F_{\textrm{p}}(x,t)$ can be separated into an equilibrium part $F_{\textrm{eq}}$ associated with the initial state and a nonequilibrum contribution $F_\textrm{p}(x,t)$ due to the periodic driving.
\begin{eqnarray}
	F_\textrm{eq}(x)&  =& 4\int_0^\infty \frac{dq}{q} \sin^2\left( \frac{qx}{2}\right) \sinh^2(\theta_q), \\
	F_\textrm{p}(x,t)&  =& 4\int_0^\infty \frac{dq}{q} \sin^2\left( \frac{qx}{2}\right)  \nonumber \\
	&\times& \left[ |\lambda_{qL}|^2 \cosh(2\theta_q) + \mathrm{Re}(\chi_{qL} \lambda_{qL}^\ast)\sinh(2\theta_q) \right] \nonumber
\end{eqnarray}
Here, $\theta_q $ denotes the Bogoliubov angles of the diagonalizing transformation for the initial equilibrium Hamiltonian obeying the equation $\tanh(2\theta_q)=V_q(1+\nu)/(2\pi v_F + V_q(1+\nu))$. In Fig.~\ref{Fig3} we show numerical results for the momentum distribution for fast and slow driving.

For $\Omega \gg \Omega^\ast$ the behavior under time evolution is consistent with the picture observed for the internal energy density in Sec.~\ref{sec:energy}. Following the initial transient dynamics the momentum distribution becomes meta-stable on intermediate times $t<t^\ast$. The momentum distribution is self-averaging in the sense that it is time-independent in contrast to the energy density where only the time average becomes constant. In the limit $\Omega \to \infty$ we recover the interaction quench limit of an effectively time-averaged Hamiltonian as for the internal energy density, see Sec.~\ref{sec:energy}. In the vicinity of the Fermi level the momentum distribution shows a nonanalytic behavior for large times
\be
	n_k(t\to\infty) - \frac{1}{2} \stackrel{\Omega \to \infty}{\longrightarrow}\mathrm{sgn}(k) |k/q_c|^{\gamma} 
\ee
with $\mathrm{sgn}(k)$ the sign function. The exponent $\gamma $ is in precise agreement with an interaction quench scenario~\cite{LL_quench}.

For slow driving $\Omega\ll\Omega^\ast$ the momentum distribution develops steps as can be seen in Fig.~\ref{Fig3}, see also Ref.~\cite{Kagan}. A similar observation has been made recently for the case of a periodic modulation of the Fermi velocity in a Luttinger liquid~\cite{Graf}. These steps may be associated with scattering processes between fermions under the absorption of quanta of the driving frequency $\Omega$. Thus, the dominant processes under the periodic driving are not only energy-conserving ones but also those where energy is conserved up to multiples of the driving frequency. Note that this step structure is remarkably similar to a simplified picture where in spirit of the work by Tien and Gordon~\cite{Tien_Gordon} for noninteracting systems the periodic driving generates a weighted superposition of equilibrium momentum distributions shifted by an energy $n\Omega$. It is, however, not possible to establish such a superposition principle precisely in the present interacting system.  Due to the parametric instability the fermions are redistributed completely for times $t\to \infty$ leading to a momentum distribution $f_k(t\to\infty)=1/2$ of infinite temperature.
\begin{figure}
	\includegraphics[width=0.9\linewidth]{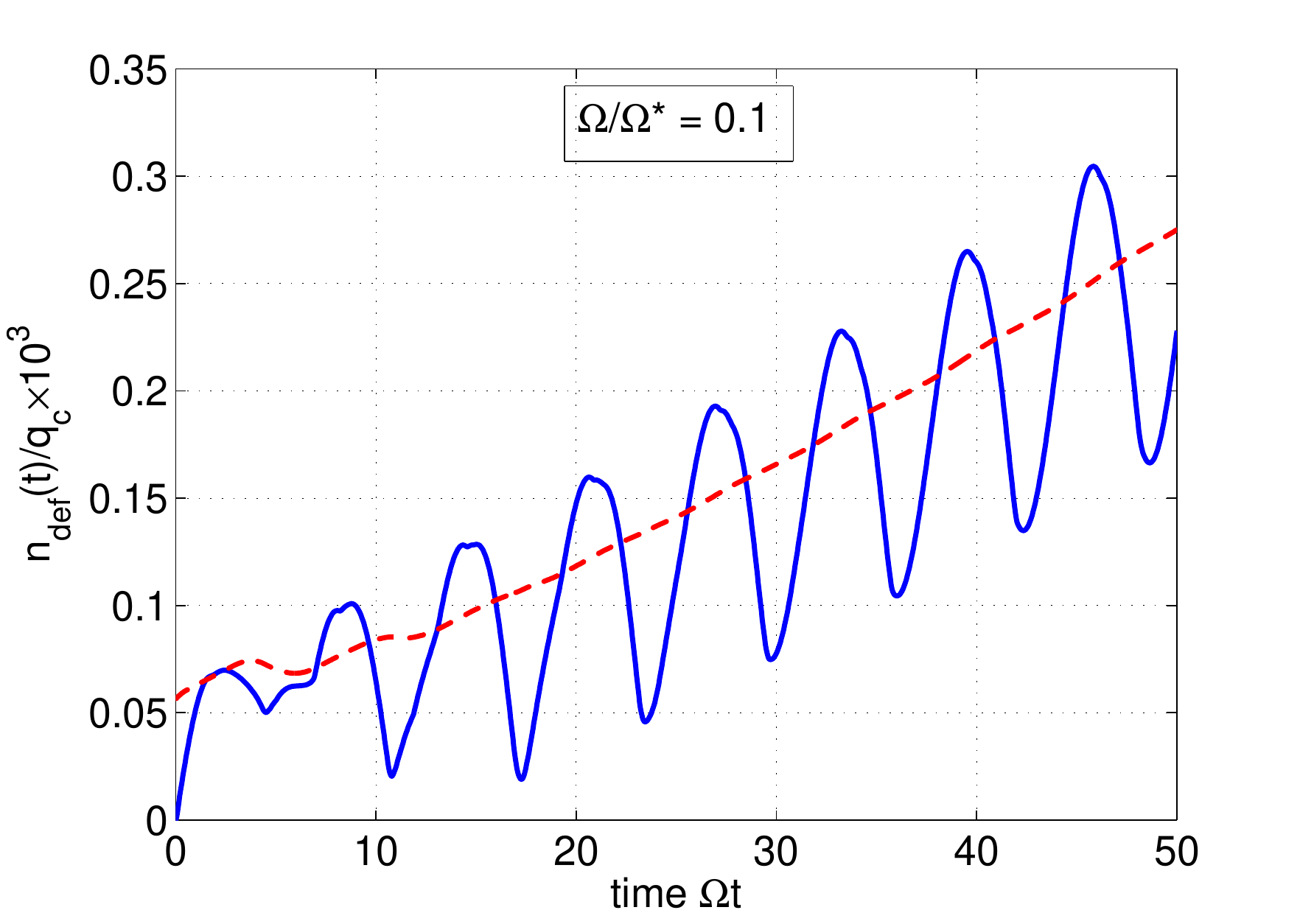}
\caption{(Color online) Time evolution of the fermionic defect density $n_\textrm{def}(t)$ (blue line) and its time average $\overline{n}_\mathrm{def}(t)$ (red dashed line) over one period. The parameters for the numerical simulations have been chosen as before.}
\label{Fig4}
\end{figure}

Due to scattering processes induced by the periodic driving fermions are redistributed in the vicinity of the Fermi surface already for small times. This excludes the possibility for an adiabatic limit where the momentum distribution follows adiabatically the external perturbation. This generates fermionic defects
\be
	n_\textrm{def}(t) = \int_0^\infty \frac{dk}{2\pi} \left[ f_k(t) - f_k^\textrm{eq}(t)\right]
\ee
compared to the adiabatic limit. Here, $f_k^\textrm{eq}(t)$ is the equilibrium momentum distribution for the instantaneous interaction potential $U_q(t)$. The integral can be restricted to positive momenta as $f_k(t) - f_k^\textrm{eq}(t)=-[f_{-k}(t) - f_{-k}^\textrm{eq}(t)]$.  As shown in Fig.~\ref{Fig4} after some transient dynamics the time-averaged defect density $\overline{n}_\textrm{def}(t)$ increases linearly in a regime where in contrast the time-averaged energy density $\overline{\mathcal{E}}(t)$ settles to a constant value on intermediate time scales which we have identified as a meta-stable state. Even though $\overline{\mathcal{E}}(t)$ approaches a constant value interaction energy is successively transferred to kinetic energy reminiscent of a heating process with an increasing entropy.

\section{Conclusions}
\label{sec:conclusions}

In this work we have studied the dynamics of a Luttinger liquid with a periodically time-dependent repulsive interaction potential. Under the periodic driving the system develops an instability due to a parametric resonance. The associated time scale $t^\star$, see Eqs.~(\ref{eq:Gamma_slow},\ref{eq:Gamma_fast}), sets the limit beyond which dissipation mechansims or internal perturbations have to be included into an appropriate description of the dynamics. On intermediate time scales before the onset of instability, it is possible to identify meta-stable states for fast and slow driving with constant time-averaged internal energies. The parametric instability generates an exponential growth of perturbations in the fermionic density leading to the buildup of a superlattice with period $x^\star$, confirm Eq.~(\ref{eq:xstar}). The fermionic momentum distribution develops a step structure that can be associated with fermiomic scattering processes under the absorption or emission of quanta of the driving frequency $\Omega$.

\acknowledgments
We thank S. Kehrein for fruitful discussions. This work was supported by the Center for Nanoscience 􏰓CeNS􏰔 Munich and the {\it Deutsche Forschungsgemeinschaft} through SFB 631 and SFB TR 12.  Financial support by the Excellence Cluster "Nanosystems Initiative Munich (NIM)" is gratefully acknowledged.

\bibliographystyle{apsrev}

\begin{thebibliography}{30}
	\bibitem{Exp_Kondo} 
		D.~Goldhaber-Gordon, H.~Shtrikman, D.~Mahalu, D.~Abusch-Magder, U.~Meirav and M.~A.~Kastner, Nature {\bf 391}, 156 (1998);
		S.~M.~Cronenwett, T.~H.~Oosterkamp and L.~P.~Kouwenhoven, Science {\bf 281}, 540 (1998);
		J.~Schmid, J.~Weis, K.~Eberl and K.~von~Klitzing, Physica B {\bf 258}, 182 (1998):
		W.~G.~van~der~Wiel, S.~De~Franceschi, T.~Fujisawa, J.~M.~Elzerman, S.~Tarucha and L.~P.~Kouwenhoven, Science {\bf 289}, 2105 (2000).
	\bibitem{Optical_lattices}
		I.~Bloch, J.~Dalibard, and W.~Zwerger, \rmp {\bf 80}, 885 (2008).
	\bibitem{Coll_quench} A.~Polkovnikov, K.~Sengupta, A.~Silva, and M.~Vengalattore, \rmp {\bf 83}, 863 (2011).
	\bibitem{AIM_per} 
		P.~Nordlander, N.~S.~Wingreen, Y.~Meir, and D.~C.~Langreth, \prb {\bf 61}, 2146 (2000);
		R.~Lopez, R.~Aguado, G.~Platero, and C.~Tejedor, \prl {\bf 81}, 4688 (1998);
		T.~K.~Ng, \prl {\bf 76}, 487 (1996);
		M.~H.~Hettler and H.~Schoeller, \prl {\bf 74}, 4907 (1995).
	\bibitem{Kondo_per}
		A.~Kaminski, Y.~V.~Nazarov, and L.~I.~Glazman, \prl {\bf 83}, 384 (1999);
		A.~Kaminski, Y.~V.~Nazarov, and L.~I.~Glazman, \prb {\bf 62}, 8154 (2000);
		Y.~Goldin and Y.~Avishai, \prl {\bf 81}, 5394 (1998);
		Y.~Goldin and Y.~Avishai, \prb {\bf 61}, 16750 (2000)l;
		A.~Schiller and S.~Hershfield, \prl {\bf 77}, 1821 (1996).
	\bibitem{Heyl_Kondo} M.~Heyl and S.~Kehrein, \prb {\bf 81}, 144301 (2010).
	\bibitem{Fal_Kim} 
		J.~K.~Freericks, V.~M.~Turkowski, and V.~Zlatic, \prl {\bf 97}, 266408 (2006);
		J.~K.~Freericks, \prb {\bf 77}, 075109 (2008);
		N.~Tsuji, T.~Oka, and H.~Aoki, \prb {\bf 78}, 235124 (2008).
	\bibitem{Fal_Kim_diss}
		N.~Tsuji, T.~Oka, and H.~Aoki, \prl {\bf 103}, 047403 (2009);
	\bibitem{Hubbard_DMFT} 
		A.~V.~Joura, J.~K.~Freericks, and Th.~Pruschke, \prl {\bf 101}, 196401 (2008);
		M.~Eckstein and P.~Werner, \prb {\bf 82}, 115115(2010);
		M.~Eckstein, T.~Oka, and P.~Werner, \prl {\bf 105}, 146404(2010);
		N.~Tsuji, T.~Oka, P.~Werner, and H.~Aoki, \prl {\bf 106}, 236401 (2011);
		M.~Eckstein and P.~Werner, \prl {\bf 107}, 186406 (2011);
	\bibitem{Hubbard_diss}
		C.~Aron, G.~Kotliar, and C.~Weber, \prl {\bf 108}, 086401 (2012);
		A.~Amaricci, C.~Weber, M.~Capone, and G.~Kotliar, arXiv:1106.3483v3 (2012).
	\bibitem{Hubbard_1D}
		T.~Oka and H.~Aoki, \prl {\bf 95}, 137601 (2005);
		T.~Oka and H.~Aoki, \prb {\bf 78}, 241104(R) (2008);
		M.~Mierzejewski and P.~Prelovsek, \prl {\bf 105}, 186405 (2010);
		M.~Mierzejewski, J.~Bonca, and P.~Prelovsek, \prl {\bf 107}, 126601 (2011).
	\bibitem{Graf} C.~D.~Graf, G.~Weick, and E.~Mariani, Europhys. Lett. {\bf 89}, 40005 (2010).
	\bibitem{Kagan} Yu.~Kagan and L.~A.~Manakova, \pra {\bf 80}, 023625 (2009).
	\bibitem{Pielawa} S.~Pielawa, \pra {\bf 83}, 013628 (2011).
	\bibitem{Holthaus} A.~Eckardt, C.~Weiss, and M.~Holthaus, \prl {\bf 95}, 260404 (2005).
	\bibitem{Poletti} D.~Poletti and C.~Kollath, \pra {\bf 84}, 013615 (2011).
	\bibitem{Schoeller_Delft} J.~von~Delft and H.~Schoeller, Ann.~Phys.~(Leipzig) {\bf 7}, 225 (1998).
	\bibitem{Landau} L.~D.~Landau and E.~M.~Lifshitz, Mechanics (Elsevier, Amsterdam, 2008).
	\bibitem{Rozhkov_Imambekov} A. V. Rozhkov, Eur. Phys. J. B {\bf 47}, 193 (2005);
		A.~Imambekov and L.~I.~Glazman, Science {\bf 323}, 228 (2009).
	\bibitem{Bettelheim} E.~Bettelheim, A.~G.~Abanov, and P.~B.~Wiegmann, \prl {\bf 97}, 246402 (2006).
	\bibitem{LL_quench} M.~A.~Cazalilla, \prl {\bf 97}, 156403 (2006);
		J.~Rentrop, D.~Schuricht and V.~Meden, New J. Phys. {\bf 14}, 075001 (2012).
	\bibitem{Tien_Gordon} P.~K.~Tien and J.~P.~Gordon, Phys.~Rev. {\bf 129}, 647 (1963).

\end{thebibliography}

\end{document}